# Home Automation System based on Intelligent Transducer Enablers

**Manuel Suárez-Albela, Paula Fraga-Lamas, Tiago M. Fernández-Caramés, Adriana Dapena and Miguel González-López**

Dpt. Electronics and Systems, Faculty of Computer Science, Universidade da Coruña, 15071, A Coruña, Spain; E-mails: m.albela@udc.es, paula.fraga@udc.es, tiago.fernandez@udc.es, adriana.dapena@udc.es, miguel.gonzalez.lopez@udc.es

* Correspondence: m.albela@udc.es; Tel.: +34 981167000 x 6051

**Abstract:** This paper presents a novel home automation system named HASITE (Home Automation System based on Intelligent Transducer Enablers), which has been specifically designed to identify and configure transducers easily and fast. These features are especially useful in situations where many transducers are deployed, since their setup becomes a cumbersome task that consumes a significant amount of time and human resources. HASITE simplifies the deployment of a home automation system by using wireless networks, and both self-configuration and self-registration protocols. Thanks to the application of such three elements, HASITE is able to add new transducers by just powering them up. According to tests performed in different realistic scenarios, a transducer is ready to be used in less than 13 seconds. Moreover, all HASITE functionalities can be accessed through an API, which also allows for the integration of third-party systems. As an example, an Android application based on the API is presented. Remote users can use it to interact with transducers by just using a regular smartphone or a tablet.

**Keywords:** home automation; IoT; plug-and-play; wireless sensor networks; transducer; self-configuration; ISO/IEC/IEEE 21451.

## 1. Introduction

The Internet of Things (IoT) paradigm refers to a global dynamic infrastructure based on interoperable communication protocols that integrate physical and virtual objects into an information network. The remarkably rapid growth of inexpensive and tiny computing devices, endowed with sensing and communication capabilities, is paving the way for the success of the IoT in almost any field. Nowadays, IoT offers a large number of opportunities for engineers, manufacturers, software developers, marketers, and investors [1,2]. Among other applications, home automation or smart homes are rapidly gaining interest.

Home automation systems are focused on improving the interaction with devices and appliances typically found indoors [3–5]. The aim of such systems is to make the home environment comfortable and accessible, automating and optimizing the use of appliances like TV sets, air conditioners, light bulbs, ovens or washing machines. For example, a home automation system might be able to modify the intensity of a light bulb through a smartphone, to monitor smoke detectors, to send alerts related to the home security system, or to automatically adjust an air conditioner to set the optimal temperature. Moreover, technologies like Bluetooth Low Energy (BLE), allow home automation systems to identify and track the presence of a specific user, making automatic adjustments to adapt the environment according to the user location and personal preferences. This interconnected ecosystem of devices would be able to offer value-added services and benefits, such as efficient energy management, environmental control, remote monitoring, reduced operational costs, security and safety.

In IoT each device must be addressed individually. Such an identification is usually carried out through an Internet Protocol address (IP address), which allows devices to be accessible over the Internet. Many standards and proprietary communications interfaces (e.g. Wi-Fi, Bluetooth, ZigBee, RFID, Z-Wave) can be used for connecting things to each other or to a cloud. The technology selected is usually determined by the physical characteristics of the environment, sensor density, desired range, and the data rates required. Among the different technologies, Wi-Fi has been presented as one of the most promising choices for future IoT developments, since it has certain characteristics commonly associated to IoT: low cost, relatively low power, compact form factor and the ability to create scalable deployments. This tendency is reflected in the approval of the IEEE 802.11s standard [6], which allows an IEEE 802.11 WLAN to be deployed as a mesh network.

Many well-established automation systems are based on wired communications, where the system must be planned in advance and installed during the physical construction of the building. For already existing buildings, such kind of wired deployments are usually more expensive, since they require to alter the internal decorations, to modify electrical wiring, and to install communication devices. Wireless solutions tackle such issues, reducing installation costs and making home automation easier to scale.

Home automation applications generally need a central control system to manage the configuration and interaction with transducers (i.e. sensors and actuators). The operation of such kind of systems poses several issues that should be solved. The main problem is related to scalability, since a single network usually connects a large number of transducers. Furthermore, all transducers would have to be able to communicate with each other and with the control system, what might derive into interoperability issues due to the heterogeneity of the sensors and actuators included. These problems become worse when the sensor network is deployed in an environment that was not originally designed for home automation.

This paper presents the design and implementation of a Home Automation System based on Intelligent Transducer Enablers (HASITE), whose transducers are automatically configured and added to the sensor network. HASITE is capable of controlling and automating house appliances through an easily manageable web interface, and presents the following features:

- The system is able to interact with each transducer of a home automation network through a scalable wireless or wired infrastructure.
- Transducers can be added to the network in a simple way. Each device is automatically registered by indicating its parameters (i.e. type, number of inputs/outputs, message format, measurement units, and refresh rate).
- The system has a simple Application Programming Interface (API) to access basic configuration parameters and the input/output of each transducer. Moreover, the access to the API can be performed ubiquitously through the Internet.
- The system has a central server that manages the sensor network and that is responsible for granting access to the API.

Thanks to the protocols implemented, the creation of new nodes and the addition of those to the system only requires the software programming of the transducers. The processes involved in obtaining the final network configuration, in detecting new nodes and in controlling such nodes, are performed automatically. Therefore, any end user is able to add a new node quickly and easily by just powering it up.

The remainder of this paper is organized as follows. Section 2 reviews the state of the art of wireless home automation communication technologies and briefly describes recent home automation systems. Section 3 introduces the most relevant plug-and-play mechanisms for transducer networks. Section 4 presents an overview of HASITE's architecture. Section 5 details the core of HASITE: the Intelligent Transducer Enablers and the plug-and-play protocol created to detect and configure the transducer nodes. Section 6 describes the API along with a usage example. Section 7 details the hardware used to implement the different subsystems, including the transducer nodes. Section 8 presents results of different tests developed to evaluate the performance of HASITE. Finally, Section 9 is devoted to conclusions.

## 2. Related Work

This section describes the most relevant communication technologies used in smart home systems. A detailed review is beyond the scope of this paper, but we recomend reference [7] to interested readers.

*2.1. Smart Home Communication Technologies*

X10 [8] was the first general purpose wired home automation network technology and remains the most widely available protocol for communicating among electronic home devices. X10 was developed in 1975 in order to provide remote control to home devices and appliances. Technologies like the Universal Powerline Bus (UPB) [9] continue the legacy of X10 to implement powerline networking, while alternatives like EnOcean [10], nanoNET [11], EIB/KNX RF [12], Z-Wave [13], 6LoWPAN [14] and Zigbee [15] avoid using powerline in favor of wireless systems.

Nowadays, the two wireless technologies that are probably most widely known are Wireless LAN (IEEE 802.11) and Bluetooth (IEEE 802.15.1). Initially, it might seem that none of them is suitable for communicating smart transducers: they are designed for transferring data at high speeds, being difficult to fit into the power constraints and cost required by sensor networks. BLE emerged to solve the problems of consumption and cost while maintaining Bluetooth range. It aims to extend the use of Bluetooth to healthcare applications, fitness or safety. One interesting feature of BLE for IoT is Internet Protocol Support Profile (IPSP), which uses IPv6 packets to discover and communicate with other devices that support the technology.

KNX has been approved as an international standard (ISO/IEC 14543-3), as well as a European standard (CENELEC EN 50090 and CEN EN 13321-1) and Chinese standard (GB/T 20965). It was developed to connect devices using twisted pair, radio waves, power line or IP/Ethernet. KNX RF uses a sub-band in the 868 MHz frequency band reserved for short-range devices (i.e. control, telemetry and alarms), supporting bidirectional and peer-to-peer transmissions.

6LoWPAN and ZigBee provide general purpose, easy-to-use, and self-organizing wireless communications for low-cost and low-power embedded devices. Both standards use IEEE 802.15.4 MAC (Medium Access Control) and PHY (PHYsical) layers, but 6LoWPAN has been explicitly designed for an easy interaction with IP-based networks.

Finally, Z-Wave is a proprietary protocol developed with an explicit focus on home control applications. Since Z-Wave uses a proprietary radio system, it suffers from fewer interoperability problems than ZigBee, but developers are attached to the technology.

Generic wireless technologies like Wi-Fi have been proposed as an alternative to specific home automation communication protocols. Thus, the Wi-Fi Alliance has recently approved a new standard for wireless connections called IEEE 802.11ah or Wi-Fi HaLow [16,17]. Wi-Fi HaLow extends Wi-Fi into the 900 MHz band, enabling the low power connectivity necessary for applications like the ones that involve the use of sensors or wearables. Wi-Fi HaLow provides robust connections in environments where the ability to penetrate walls and other obstacles is an important requirement. In addition, Wi-Fi HaLow is expected to be backward compatible with Wi-Fi and deliver other benefits, such as multi-vendor interoperability, strong government-grade security, and easy setup.

Regarding interoperability, the IEEE 1905.1 standard has been recently developed for enabling the convergence of the heterogeneous technologies typically found in digital home networks [18].

Such a standard provides an interface for common home network technologies, allowing for the coexistence of data link and physical layer protocols such as IEEE 1901 over power lines, Wi-Fi, Ethernet over twisted pair or fiber, and MoCA 1.1 over coaxial cables [19].

Cellular communication technologies defined by 3GPP [20], 4G machine-to-machine communications [21] or MTC [22] are still emerging. For instance, in [23] the authors analyze the potential of 5G cellular systems for being a key driver in a global IoT, considering both the technological and standardization aspects. The paper also illustrates the massive business shifts that a tight link between IoT and 5G may cause in the operator and vendor ecosystems.

Table 1 provides an overview of common features of wireless protocols: frequency band, data rate, and maximum number of nodes. All technologies supporting the 868 MHz frequency band, except KNX RF, also support operating at 908 MHz.

Table 1. Comparison of relevant home automation communication protocols.

|  | **Frequency band** | **Data rate** | **Max. nodes** | **Wired/Wireless** |
|---|---|---|---|---|
| UPB | 4-40KHz | 480 bits/s | 64, 000 | Wired |
| X10 | 433 MHz (EU) | 20 bit/s | 256 | Wired |
| EIB/KNX RF | 868 MHz | 16.4 kbit/s | 256 per line | Wireless |
| Z-Wave | 868 MHz (EU) | 9.6 kbit/s | 232 | Wireless |
| nanoNET | 2.4 GHz | 2 Mbit/s | $2^{48}$ | Wireless |
| EnOcean | 868 MHz (EU) | 120 kbit/s | $2^{32}$ | Wireless |
| IEEE 802.15.4/ZigBee | 868 MHz (EU), 2.4 GHz | 20-250 kbit/s | 65, 536 | Wireless |
| IEEE 802.15.4/6LoWPAN | 868/915 MHz, 2.4 GHz | 100-250 kbit/s | - | Wireless |
| IEEE 802.11ah/Wi-Fi HaLow | 900 MHz | 18 Mbit/s | - | Wireless |

*2.2. Home Automation Systems*

In the last years, several relevant home automation systems have been proposed. For instance, in [24] the authors designed, implemented and tested a method for integrating different wireless technologies (ZigBee, Wi-Fi, GSM/GPRS) in heterogeneous home automation environments. This LabView-based system uses wireless nodes that are distributed within an intelligent building, monitoring different ambient parameters (temperature, humidity, light) and controlling different subsystems (e.g. lighting or irrigation). ZigBee transceivers have been also used in [25], where the authors have oriented the architecture towards the construction of a home automation system based on intelligent power outlets.

A complete remote security platform is proposed in [26]. The solution integrates several home automation technologies and uses EIB/KNX RF and ZigBee to communicate with in-home appliances. The authors also propose an IP-based communication protocol between the home automation controller and the rest of the equipment. The platform can be controlled by remote security providers and homeowners through the Internet, via secure and efficient communication protocols.

In [27] the authors propose a home network formed by sensors and actuators that control and obtain information from the home environment. The user interface is a low-cost Android tablet, while the controller is an Arduino board that offers connectivity with the electronic devices of the home network. The system uses both Zigbee and X10: Zigbee ensures the reliability of the system, while X10 allows for the integration of low cost controllers to optimize the overall cost of the system.

Android is at the core of the system presented in [28]. A smart home application communicates mobile devices with a micro web-server via the Internet using a REpresentational State Transfer (REST) based web service. Wi-Fi or a 3G/4G connection can be used to connect the user device to the Internet. The system integrates elements such as light switches, power plugs and temperature sensors. The authors tested the system considering several home operations (e.g. remote switching, automatic environmental control or intrusion detection).

The authors in [29] present CONDE, proposing a decentralized approach for decision and control for smart building applications. Compared to traditional centralized systems, response time and power consumption are both reduced. The tests performed using HVAC and lighting systems obtained lower decision delay times and better energy saving values than other alternatives compared.

Finally, it is worth mentioning the paper in [30], which analyzes the evolution of home automation systems towards IoT. The authors conclude that no commercial solution based on such kind of systems is easily applicable to the concept of IoT. Therefore, currently it is necessary to translate the different protocols to IP, with the extra cost and overhead that such a process entails. Furthermore, it is emphasized that home automation technologies usually make use of dedicated and proprietary communication buses, requiring the use of proprietary end-nodes (i.e. sensors, actuators) and protocols. This is the case of technologies like Building Automation and Control Networks (BACnet), European Installation Bus (EIB) and Local Operating Network (LonWorks). Moreover, these technologies are not interoperable at all, so the user is limited to the devices available for each technology.

## 3. Plug-and-Play Mechanisms for Transducer Networks

Home automation systems can benefit from the plug-and-play connection of transducers to the networks nodes. It allows for simplifying application development, network deployment, service procedures, and on-the-fly substitution of the sensors. A transducer can be considered plug-and-play when it becomes fully operational and networked after powering it up. This plug-and-play feature provides significant benefits:

- Application development, sensor network deployment, and services are simplified noticeably.
- Wireless Sensor Networks (WSN) nodes can be upgraded or reassigned dynamically for new tasks by just changing sensors.
- Sensors that become disconnected from a node due to unexpected reasons can be automatically taken out of use.

It is important to emphasize that, since it has not been established a common standard, there is not a straight answer to certain questions on the development of the system, such as how to communicate with the transducer network (wired or wireless), which set of protocols should be used to manage all the elements, how to perform the system configuration, or which are the rules to follow to control the system automatically.

### 3.1. The ISO/IEC/IEEE 21451 standard

The most relevant initiative for adding plug-and-play capabilities to intelligent transducers is the ISO/IEC/IEEE 21451 standard [31,32] (previously known as IEEE 1451). This standard constituted the basis and inspiration when designing HASITE's architecture.

The ISO/IEC/IEEE 21451 standard is aimed at providing sensors and actuators with a series of features like self-identification, self-description, self-calibration and standard data formats. These mechanisms facilitate interoperability between networks with heterogeneous technologies.

The Transducer Electronic Data Sheets (TEDS) are one of the most important components of any system based on ISO/IEC/IEEE 21451. TEDS are similar to the traditional data sheets provided by hardware suppliers, containing information such as the manufacturer's name, the type of transducer, the serial number, the measurement range, the accuracy, the calibration data, and the supported data formats. The standard also supports the implementation of new kinds of TEDS for proprietary solutions.

The complete ISO/IEC/IEEE 21451 family of standards defines a set of common, open, network-independent communication interfaces for connecting smart transducers to digital systems and instruments in network-independent environments. ISO/IEC/IEEE 21451.0 defines a set of common functionalities, TEDS and commands for accessing the transducers connected in various physical configurations (e.g. point-to-point, distributed multi-drop, and wireless configurations) to fulfill different application requirements. The standard family presents three possible ways for accessing sensors and actuators from a network: IEEE 21451.1, IEEE 21451.0 Hyper Text Transfer Protocol (HTTP) and the Smart Transducer Web Services [33,34].

Figure 1 shows the architecture of an ISO/IEC/IEEE 21451 system. It is divided into two subcomponents: the Network Capable Application Processor (NCAP) and the Transducer Interface Module (TIM).

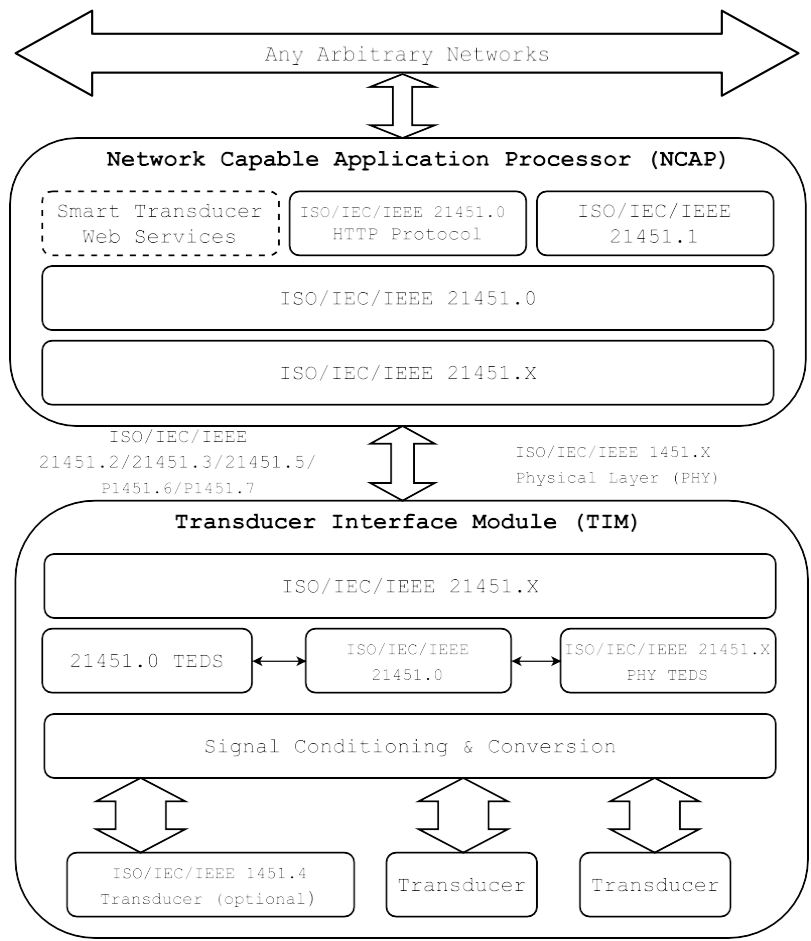

**Figure 1.** ISO/IEC/IEEE 21451 architecture.

The NCAP is a special network node that performs processing and communication tasks. Its radio interface allows it to communicate with intelligent transducers while offering different web services through their interfaces. The NCAP may run an HTTP server to provide the sensor network with access to each of the transducers via the Internet. The NCAP has a network interface that provides the services offered by the transducers. When a request is received, the NCAP uses a service to send the necessary commands to the transducer module through the communications module. Then, the transducer module processes the request and sends back the response to the NCAP. Finally, the NCAP uses its HTTP server to communicate the response to the remote client.

The Intelligent Transducer Module, referred to as TIM, Smart TIM (STIM) or Wireless TIM (WTIM) in the standards, acquires sensor signals and controls the actuators. It is able to provide sensor data or actuate over remote elements through services. The communication with the rest of the network and with remote users is conducted through an embedded air interface. The TIM is divided into two interfaces: the Network Interface (NI) (a network communications protocol for connecting the NCAP to the network) and the Transducer Independent Interface (TII) (a protocol for transferring the information about the transducer).

The ISO/IEC/IEEE 21451 standard specifies the basis for interoperating and scaling heterogeneous sensor networks. Even though it provides a high degree of flexibility, it also requires to carry out a complex deployment and implementation. Such a complexity and the cost of meeting all functionality often involves the use of a separate microcontroller or a processor in the TIM, leading to a significant increase in both cost and power consumption. This issue constitutes a problem, especially in the case of creating scalable transducer networks. For these reasons (among others), as of writing, the standard has not been very successful in terms of functional implementations [35,36].

One example of implementation is carried out in [37]. There, the author describes a plug-and-play connection to WSN nodes for ISO/IEC/IEEE 21451 smart sensors. The system includes multiple useful features, but it is rather complicated, what leads to expensive devices. Another example can be found in [38], where a system for evaluating seawater quality is presented. The authors conclude that the additional complexity required by ISO/IEC/IEEE 21451 command packet recognition increases the communications lag by 50 %. The authors explain that, although such a delay may appear excessive for specific cases, it can be considered negligible when using modern microcontrollers.

*3.2. Alternatives to the ISO/IEC/IEEE 21451 standard*

There are not many alternatives to the ISO/IEC/IEEE 21451 standard. Just a few articles can be found in the literature about systems that describe reconfiguration capabilities for WSN nodes. All of them are based on proprietary protocols.

An example is [36], where a novel TEDS template is introduced, implemented and evaluated using a Programmable System on Chip (PSOC). The system proposed simplifies both the management and the integration of new transducers. The reconfiguration can be performed on-the-fly, either based on an algorithm, or by an NCAP request.

Other researchers describe alternative systems oriented towards simplicity. For instance, in [39] it is proposed a framework that uses a plug-and-play mechanism to create groups of transducers in order to minimize network power consumption. Another example is described in [40], where a simplified plug-and-play system is presented. Such a system is able to detect the connection and disconnection of a Commercial Off-The-Shelf (COTS) sensor, to identify the connected sensors and to recover the necessary data from the sensor network. Nevertheless, this approach lacks the capability of unique ID generation, what might lead to a scenario where two or more transducers could be identified as the same one.

The present article deepens into these characteristics and, departing from the basis of the ISO/IEC/IEEE 21451 standard, seeks to remove most of its complexity.

**4. Home Automation System based on Intelligent Transducer Enablers**

As it was described in Section 3, the main plug-and-play mechanisms for transducer networks are aimed at making the transducers accessible without having to configure the network they are connected to. This approach, despite being useful in some scenarios, has two main problems. The most obvious one is that it requires to implement the behavior of several hardware elements and their communication protocols. This increases the total cost and the complexity of the overall system. Currently, the ISO/IEC/IEEE 21451 standard has been available for years, but, as far as we are aware, it has not been implemented in any successful commercial project that required a large deployment. The second problem is that the standard it is not oriented towards IoT. Even though it supports HTTP communications, the whole idea of TEDS and how transducers and their data are defined, was not conceived with the Semantic Web in mind. This would require an additional layer on the standard, increasing the current overhead that it already imposes.

HASITE (Home Automation System based on Intelligent Transducer Enablers) has been designed for tackling the two issues mentioned. On the one hand, it has been optimized for easing intelligent transducer node identification and configuration within a smart home deployment. On the other hand, HASITE provides easy, understandable and parseable access to the transducers and their data. Thus, HASITE is characterized by its ability for self-configuring and auto-describing transducers nodes, and for using wireless communications.

*4.1. Hardware Architecture*

Figure 2 illustrates the main elements present in HASITE's hardware architecture, which can be separated into two groups:

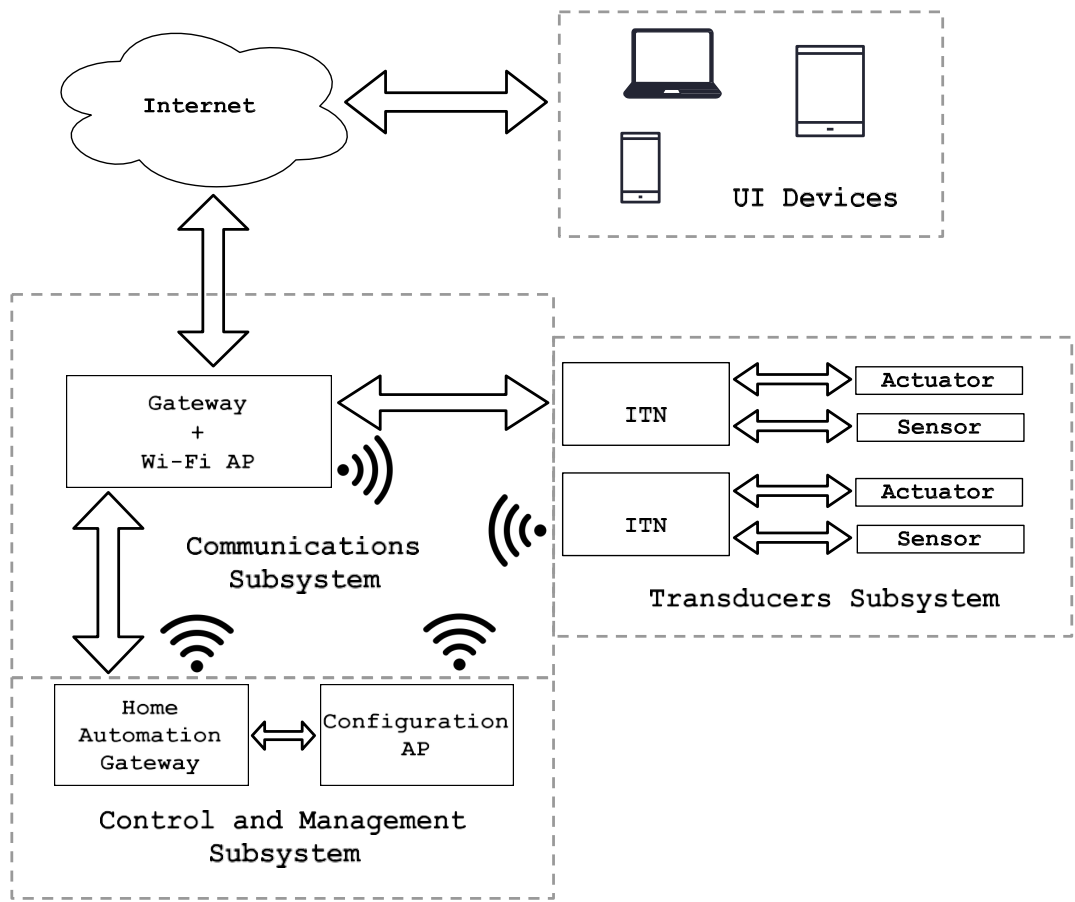

**Figure 2.** HASITE's architecture.

- HASITE-independent infrastructure: elements already available at most homes but needed by HASITE to achieve full functionality. Such elements are also used for other purposes:
  - Internet.
  - Gateway: it provides Internet access.
  - Network access point: it connects the nodes to the Home Automation Gateway.
  - User Interface Devices: smartphones, tablets or personal computers that allow for the interaction with HASITE.
- HASITE-dependent infrastructure. It includes the hardware elements directly used by HASITE:
  - Home Automation Gateway: it registers and provides access to the nodes. It also provides external access from the Internet through the Gateway to facilitate remote control and monitoring. It is connected to a configuration Access Point to support the self-configuration mechanism. It also supports and grants access to the API.
- Configuration AP: it is a special Access Point used for the auto-configuration protocol. It can be a traditional AP or a transducer node of the network with AP capabilities.
- Intelligent Transducer Node (ITN): it is a node consisting of a number of transducers, a processing system, and wired or wireless communication modules that allow for exchanging data with the Home Automation Gateway. Figure 3 shows a block diagram of an ITN. Note that ITNs within a home automation network can control sensors (that send environmental information) and actuators (that perform specific tasks). For instance, temperature, humidity, light, presence and electrical current sensors were the basis of different ITNs built during the development for HASITE. Moreover, actuator-based ITNs included relays, which switch power sockets on and off, and light intensity controllers.

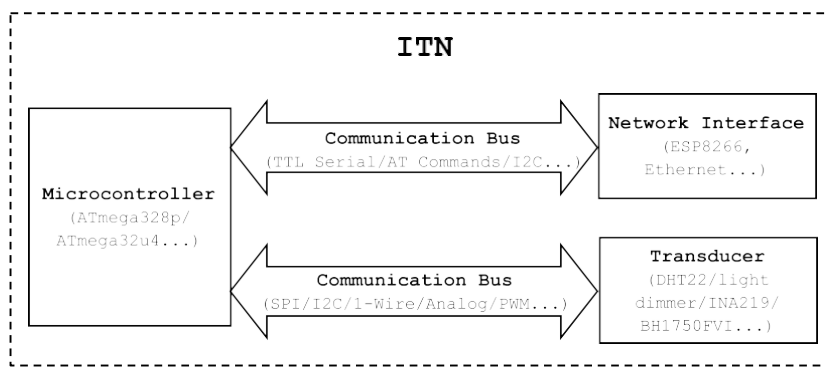

**Figure 3.** Generic ITN components and communication buses.

It is important to note that the Home Automation Gateway functionalities can be carried out by any regular PC [26,41,42], but such an element requires dedicated hardware with high dimensions and consumption. For the experiments detailed in Section 8 we opted for a Single Board Computer (SBC), like other authors chose in recent publications [27,28]. An SBC is a small board with one or several microprocessors, memory, input/output (I/O) pins, and other features typically found in a traditional computer. The use of an SBC is very attractive for home automation because its low cost, small size, and low consumption.

In contrast to this approach, a decentralized system [29] could yield better power consumption (a central processing element is not required) and lower response times for automatic decision making. However, such a system would increase the complexity of HASITE, mainly the plug-and-play protocols and the API access. Thus, using a central control element was preferred instead.

*4.2. Software Architecture*

The software core of HASITE's architecture is called Self-Configuring Intelligent Transducer System (SCITS) (detailed later in Section 5). SCITS involves two main elements: the ITEs (described in Subsection 5.1) and the self-configuration and detection protocols (detailed in Subsection 5.2). In Figure 2 the different subsystems needed to manage such elements can be observed:

- Communications subsystem. This subsystem interconnects all transducers and provides Internet access.
- Control and management subsystem. This subsystem is responsible for controlling and managing the rest of the subsystems. It monitors and registers all the ITNs, it obtains data from the sensors at a proper frequency, and controls the actuators according to the user preferences. It also grants and controls the access to the API.
- Transducers subsystem. It consists of groups of ITNs. Each ITN may contain one or more transducers.

It is also important to note that integral control and communications are key in home automation. For such a purpose, the Home Automation Gateway receives and processes the data acquired by the sensors while monitoring and preserving the connection between all elements. It also provides the persistence and external access needed for user interaction with the whole system. In particular, the most important functionalities of the Home Automation Gateway are the following:

- Central management of ITNs. It registers and keeps control of the ITEs available on the system, it manages the communication with the ITNs, and it processes the requests from other modules. It is also the responsible for coordinating the self-configuration and self-registration protocols.
- System configuration. It guarantees the persistence of critical data, like the information about the ITEs, the system configuration, or the data about the users of the system.
- Gateway. It provides network connectivity.
- Software interface. It contains the Application Programming Interface (API) that allows third party applications to access all the system functionalities.

## 5. HASITE: Self-Configuring Intelligent Transducer System

As it was mentioned in Section 4, SCITS is the core of HASITE. The most important elements of SCITS are the Intelligent Transducer Enablers (ITEs). Such elements are similar to ISO/IEC/IEEE 21451 TEDS, but they work at a node level instead of at a transducer level. An ITE describes an ITN and all the transducers associated to it. The main objective of the SCITS system is to make the ITNs available to the network with minimum user interaction in the smallest amount of time. This is essentially achieved thanks to the self-configuration and self-registration mechanisms.

*5.1. Intelligent Transducer Enablers*

On the left of Figure 4 it is depicted a representation of all the ITNs included in HASITE at a given moment. They are represented as a response from the API. We chose JSON to represent the information due to its simplicity and parsing capabilities. In order to build an ontology around the ITNs and ITEs, other notations could be used, such as XML or RDF [43] (e.g. JSON-LD, Turtle).

On the right side of Figure 4 the complete ITE information of the first ITN listed can be seen. This information is related to a light regulator ITN. In this case, only the essential information for transducer identification and communication is present. Additional information such as calibration parameters or physical ITN location could be included.

The first fields of the data include parameters that univocally define the ITN, like the database internal ID, the serial number or the IP currently assigned to the ITN. Next, the ITE of the transducer is represented, including a descriptive name, the ITE type and version, and the list of sensors and actuators present. In the case shown the ITN only has one actuator, the one that allows the regulation of the light intensity. For such an actuator four parameters are defined: a descriptive name, a Uniform Resource Identifier (URI) of the transducer, and two additional fields. The first additional field ("json_req") includes the data related to an ITN that can be read or write by using PUT or GET requests. In this case, the request must include a JSON object with a field called "ActuatorValue". The data type of this field is "Unsigned Int" and it supports a range from "0" to "100" with a resolution of "1.0". The second field ("json_res") presents the message format of the response that an ITN will send when a PUT request is issued. A Boolean field will indicate whether the request was successfully processed or not (in case the message was malformed or the values were out of the specified range).

ITEs, like TEDS, can be stored locally (i.e. on the ITN) or remotely (i.e. on the Internet, in a machine in the local area network, or even in the Home Automation Gateway). In HASITE's implementation, each ITN stores only a unique identifier, formed by the ITE identifier and the ITN serial number. This unique identifier is used by the Home Automation Gateway to fetch the ITE when the self-configuring and self-registration mechanisms take place, or when the information about an ITN is accessed through the API. For instance, in the implementation whose results are presented in Section 8, the unique identifier is a group of three numbers that follow the structure shown in Table

```
1  [
2    {
3      "id": 8,
4      "sn": 1,
5      "ite": {
6        "name": "Light Regulator 220VAC",
7        "type": 6
8      }
9    },
10   {
11     "id": 11,
12     "sn": 1,
13     "ite": {
14       "name": "Temperature and humidity DHT-22",
15       "type": 2
16     }
17   },
18   {
19     "id": 12,
20     "sn": 1,
21     "ite": {
22       "name": "Switch",
23       "type": 7
24     }
25   },
26   {
27     "id": 13,
28     "sn": 1,
29     "ite": {
30       "name": "Power Meter INA",
31       "type": 10
32     }
33   },
34   {
35     "id": 24,
36     "sn": 2,
37     "ite": {
38       "name": "Temperature and humidity DHT-22",
39       "type": 2
40     }
41   }
42 ]
```

```
1  {
2    "id": 8,
3    "sn": 1,
4    "ip": "192.168.1.144",
5    "ite": {
6      "name": "Light Regulator 220VAC",
7      "type": 6,
8      "version": 1,
9      "sensors": [],
10     "actuators": [
11       {
12         "name": "Light Regulator",
13         "json_req": {
14           "name": "ActuatorValue",
15           "units": "Light intensity percentage",
16           "data_type": "Unsigned Int",
17           "min_value": "0.000",
18           "max_value": "100.000",
19           "resolution": "1.000"
20         },
21         "json_res": {
22           "name": "ActuatorSet",
23           "units": "-",
24           "data_type": "Boolean",
25           "min_value": "0.000",
26           "max_value": "1.000",
27           "resolution": "0.000"
28         },
29         "uri": "/actuators/0"
30       }
31     ]
32   }
33 }
```

**Figure 4.** List of the transducers included on HASITE (left) with a detailed ITE description of the light regulation ITN (right).

2. Using this type of identifier, there are 16 777 216 types of ITN, 4 294 967 296 serial numbers and 256 possible revisions of an ITN. A more complex identifier could be created and easily obtained by using the unique MAC address of the network interfaces of the ITN.

The following are the most important benefits of using ITEs:

- ï An ITE enables transducer self-identification. Thanks to the information carried by an ITE every transducer can be identified and described by the host.
- ï An ITE provides long-term self-documentation. For instance, the ITE of a sensor can be updated to store information such as the location of the sensor, its re-calibration data, repair records, or maintenance-related data.
- ï The automatic transfer of ITE data to the network avoids entering sensor parameter values by hand, what might be affected by human errors.
- ï An ITE eases the maintenance, upgrade, and field deployment of the sensors. This helps to reduce the total life-cycle cost of system.

**Table 2.** Memory division of the microcontroller EEPROM in Bytes

| Bytes | 0-2 | 3-6 | 7 | 8 | 0-40 | 41-104 | 105-108 | 109-124 |
|---|---|---|---|---|---|---|---|---|
| **Data** | Sensor type | SN | Version | Status | SSID | Password | IP Gateway | IPv6 Reserved |

- The plug-and-play capability offered by an ITE allows the nodes to be connected with other standardized physical communications media and to keep on operating without requiring any software change.
- The transducer information and its messages are auto-descriptive. They are also parseable, making the interaction with third party systems systematic.
- The ITEs define the transducers in a way they can be used by the Semantic Web. They can be easily represented by using JSON-LD, allowing for the use of Resource Description Framework (RDF) and enabling the creation of an ontology around transducers and their relationships. Each transducer can be seen as a resource that can be addressed through a unique IPv6 (assigned to its ITN). This would grant access to any transducer present on the node a World Wide Web URL without any further configuration.

*5.2. Self-configuration and self-registration protocols*

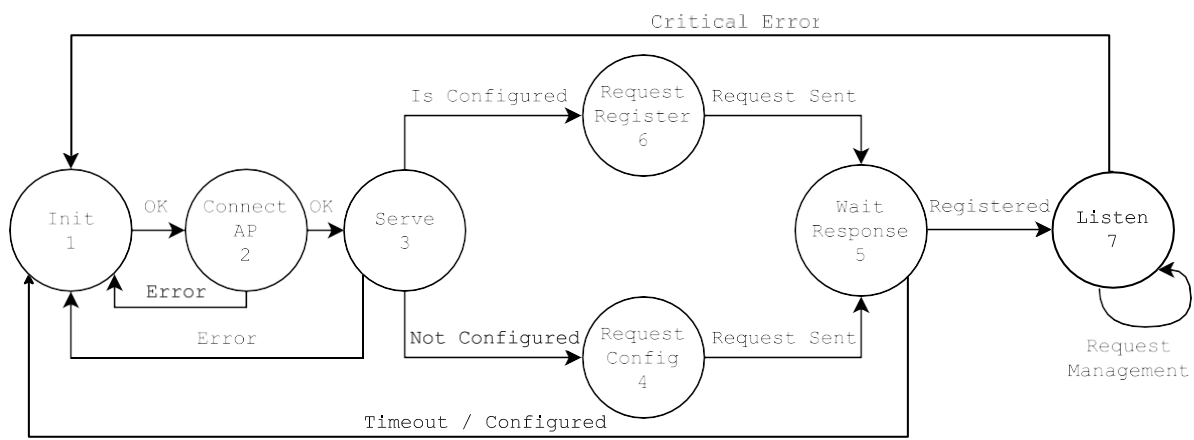

**Figure 5.** Finite State Machine implemented by an ITN.

The behavior of each ITN is modeled by the Finite State Machine (FSM) represented in Figure 5. From the moment an ITN is powered up, until it is ready to be accessed, it runs through the following states (note that the states named *X-A* represent the self-configuration phase, while the ones named *X-B* are related to the self-registration phase).

- Self-configuration phase:

    - 1-A Init: configuration of default values. The microcontroller reads from its internal memory values like the SSID and the password of the network it will connect to, and starts the FSM.
    - 2-A Connect AP: the connection to the configuration router is established.
    - 3-A Serve: the TCP server is started on the ITN to listen for incoming requests.
    - 4 Request Config: the configuration request is issued and sent to the Home Automation Gateway.
    - 5-A Wait Response: the ITN awaits for a response that contains the configuration parameters. After receiving the configuration, the FSM writes it to the non-volatile memory of the ITN and restarts the ITN.

- Self-registration phase:

    - 1-B Init: configuration of default values. In this state the microcontroller reads the configuration previously received.
    - 2-B Connect AP: the connection to the default router is established.
    - 3-B Serve: the TCP server is started on the ITN to listen for incoming requests.
    - 6 Request Register: the ITN issues a register request and sends it to the Home Automation Gateway.
    - 5-B Wait Response: after the Home Automation Gateway registers the node, it sends an ACK message. In this moment the ITN becomes available to the rest of the network.
    - 7 Listen: the ITN remains listening for incoming requests to serve and handle the transducer processing and communication operations.

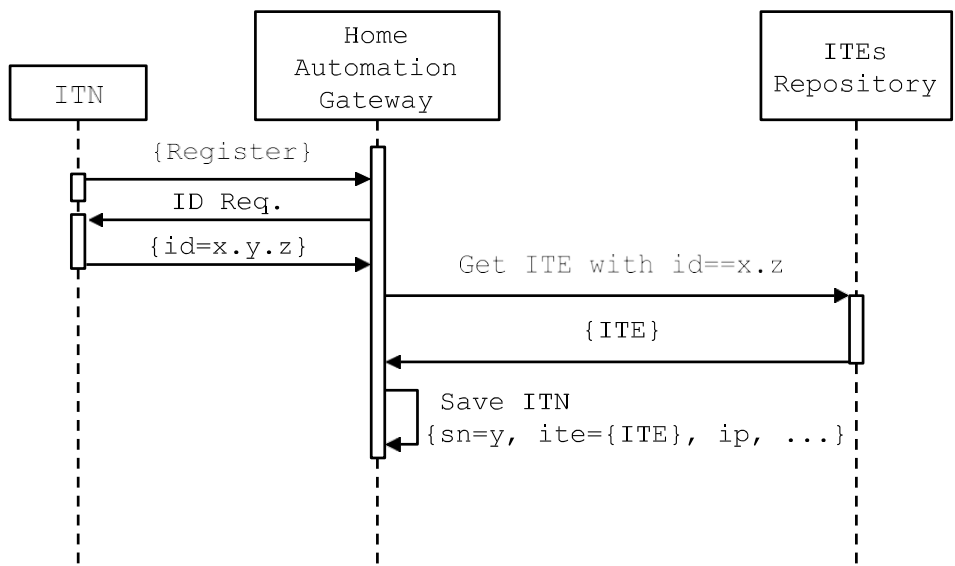

**Figure 6.** Self-registration protocol sequence diagram.

After states 4 and 6, the Home Automation Gateway receives specific requests that contain the ITN identifier. Before sending a response, the Home Automation Gateway queries its repository of local and/or remote ITEs searching for a match between the ITN identifier and the corresponding ITE. In case of finding a match, it registers the ITN in conjunction with its IP and a timestamp. This process is illustrated in the sequence diagram shown in Figure 6, which represents the exchange of messages between an ITN, the Home Automation Gateway, and the ITEs repository.

When the ITN is in state 7, the API is able to list it, allowing remote users to obtain all the information about the node (its IP and the information associated with the ITE) and to carry out modifications on certain fields (e.g. descriptive name, location, calibration), as well as to interact with the ITN transducers.

Using the API, any authorized application can get data from the sensors or perform actions on the actuators of the ITN. It is important to note that the FSM previously described has been simplified for the sake of clarity, meaning that any ITN connected to the network could be added to the system without any confirmation. More secure options are supported, being only necessary to configure the system for a required confirmation before adding a node to the network.

Two additional mechanisms are defined by using the ITN unique identifier (ID):

- Dynamic request. When the Home Automation Gateway receives a configuration or a register request message, it sends a message to every client that is using the API. If there are no clients connected, the request is stored and sent in the next connection. The message contains the unique ID of the ITN, so that the user can easily verify that the request corresponds to the specific ITN being added. Finally, when the user acknowledgment is received, the process is executed normally. This mechanism, although simple, provides an additional security level.
ï Previous registration of identifiers. If a large number of ITNs is going to be deployed, they can be previously added to the system through a list of IDs. In the self-configuration process, when the Home Automation Gateway receives a configuration request message, it checks whether the ID of the ITN that issues the request is present or not in the list of known devices. If the ID is present, the system proceeds normally. In any other case, a confirmation message is sent to the available users.

*5.3. Comparison with ISO/IEC/IEEE 21451 standard*

Finally, in order to clarify the differences between HASITE's architecture and the one proposed by the standard ISO/IEC/IEEE 21451, Figure 7 compares both.

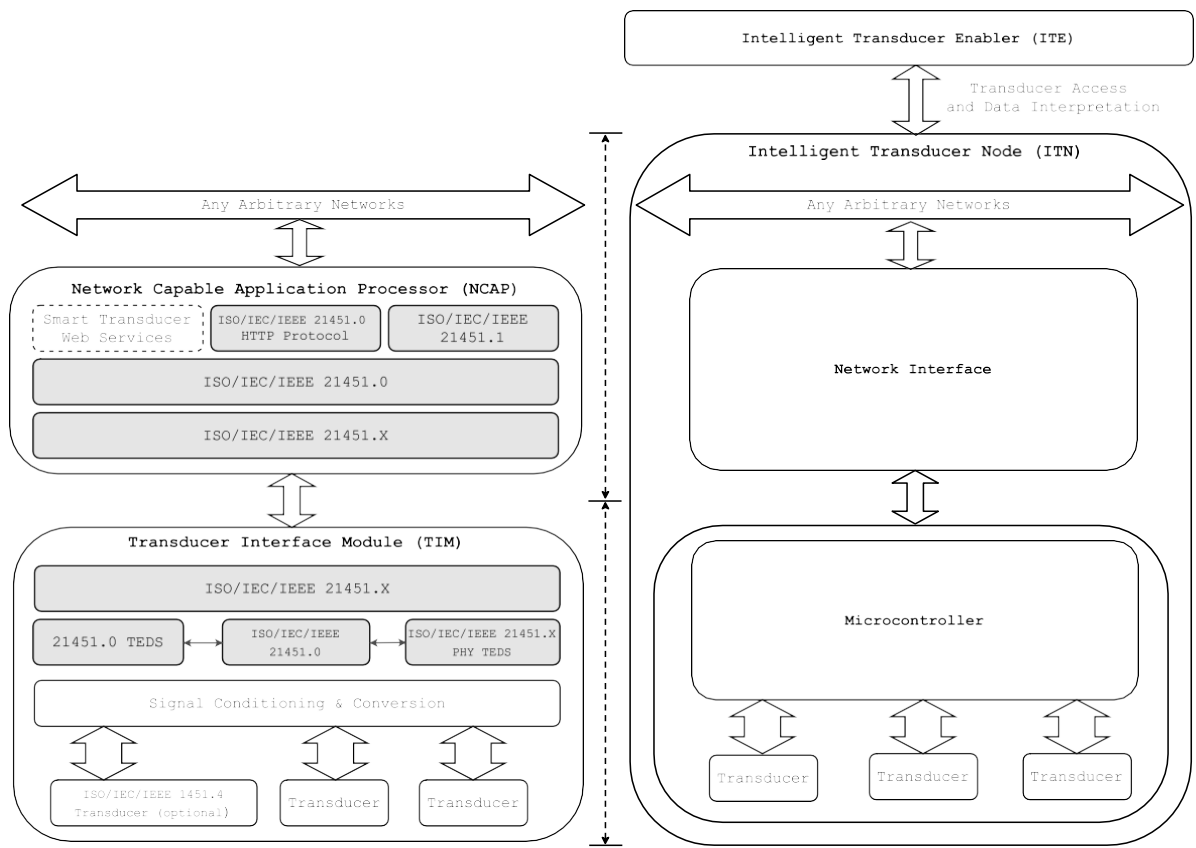

**Figure 7.** Comparison of ISO/IEC/IEEE 21451 and HASITE architecture.

As Figure 7 reflects (and as explained in Section 3), the ISO/IEC/IEEE 21451 standard defines several elements (i.e. TIM, NCAP, Transducers, Network) and specifies the communication protocols among them. In such a standard approach, transducers, TIM and NCAP demand processing capabilities higher than the ones actually required. Moreover, the communication protocols between the elements introduce overhead to the data transferred through the network. Therefore, these protocols add complexity to the system and may reduce the number and type of compliant transducers.

In contrast, HASITE allows for using any COTS transducer, concentrating the intelligence into a single element (into the ITN). The communications between the Network Interface, the Microcontroller and the Transducers are not constrained in any way. An ITN only needs to implement the self-configuration and self-registration protocols, and to have an ITE to allow access to its transducers.

With this approach we remove the dependency on the transducers to be compliant with a standard. This is related to one of the main limitations of ISO/IEC/IEEE 21451: each transducer has to implement the whole standard, what derives into higher development costs (i.e. transducers require extra hardware and complexity to implement their TEDS). Moreover, in an ISO/IEC/IEEE 21451 system, TIM and NCAP also have to be implemented and compliant with the specifications. Therefore, the whole system makes a complete implementation not only complex but expensive, with a cost that increases with the number of transducers.

## 6. HASITE: API and user interface

HASITE implements an API that allows third-party applications to access the main features of the system. For the examples below, it is assumed that the API of HASITE is listening on the IP address 172.16.0.2 and port 5050. The major API endpoints allow remote users to:

- List all the registered ITNs. This is accomplished by issuing a GET request to "http://172.16.0.2:5050/transducers" (the response is represented on the left in Figure 4).
- Access the basic information of an ITN, like its IP or the associated ITEs. With the "id" field obtained from the previous API response (i.e. the light regulator, represented with an "id" 8), all the ITN information can be obtained by issuing a GET request to "http://172.16.0.2:5050/transducers/8" (this information conforms the ITE, as shown on the right of Figure 4).
- Execute actions like reading a sensor value or changing the state of an actuator. Again, from the previous response, we can obtain the actual intensity of the light regulator transducer by issuing a GET request to "http://172.16.0.2:5050/transducers/8/actuators/0", obtaining as a result a JSON response "{"ActuatorValue" : 50}". Note that the "actuators/0" part is obtained from the field "uri" of the ITE. As it is explained in Section 5.1, this transducer values range from 0 to 100, so it is easy to understand that the regulator controlled by this ITN is now sending half of the power to any light emitter connected to it. To change the value of the actuator, and using again the information obtained from the ITE, it is only necessary to issue a PUT request to "http://172.16.0.2:5050/transducers/8/actuators/0" with a JSON message that indicates the new value: "{ "ActuatorValue": 20 }". If the request is processed successfully, the response will be "{"ActuatorSet": 1}". In any other case, the response will be "{"ActuatorSet": 0}", indicating that it was not possible to change the actuator state.

The API and the persistence of the different system elements were built on top of the Django framework [44], which is a free and open source web application framework written in Python. It consists of an object-relational mapper (ORM) (that mediates between data models and a relational database), a system for processing Hypertext Transfer Protocol (HTTP) requests with a web template system, and a regular-expression-based Uniform Resource Locator (URL) dispatcher. The whole software solution developed is light enough to run in a Raspberry Pi 2. The ORM is used to persist the ITEs, the user and system configuration, and the registered ITNs data. The system also supports logging data samples associated with a particular ITN.

The access to the API endpoints is secured with HTTP Basic Authentication, but more complex and secure mechanisms can be used. For instance, the API of the system evaluated in Section 8 is based on Django serialization capabilities, which support advanced authentication protocols like OAuth 2.0.

In order to illustrate better the API capability for developing graphic user interfaces, a hybrid smartphone application for Android and iOS was implemented. Using the smartphone Bluetooth interface, an indoor location system was developed. The application, as it can be seen in Figure 8, allows us to create relationships between BLE beacons and ITNs. BLE beacons are Bluetooth Low Energy devices that periodically emit a message with some parameters that allow for unique identification. Smartphones can read these messages and use the Received Signal Strength Indicator (RSSI) to estimate the distance from the beacons. This allows for creating a simple but effective indoor location system.

The application can run in two different modes: default and BLE. In default mode (see A in Figure 8) the application displays all the ITNs of the system (i.e. a light regulator and a temperature sensor). In BLE mode (B and C in Figure 8), when a beacon is detected, only the ITNs associated with the beacon are shown, what creates a better user experience. For example, in Figure 8 it is illustrated the behavior of the system when a light dimming control and a temperature sensor are paired with two different BLE beacons. When the user gets closer to one BLE beacon the smartphone detects it and shows just the light dimming control (B) or the temperature sensor (C) depending on which one is closer, what eases the interaction with the transducers depending on the user location. Also note the list of numbers presented below the ITN in (B) and (C), which are the RSSI values of every BLE beacon detected in the environment (they are only displayed for debugging purposes).

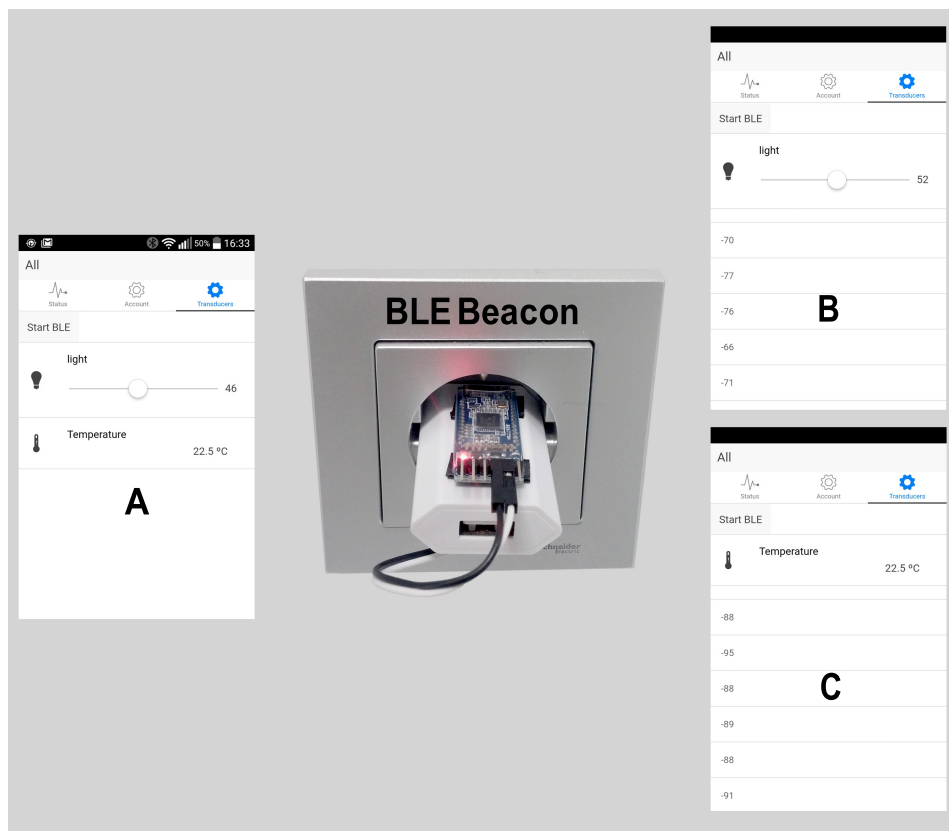

**Figure 8.** Android application in default (A) and BLE (B and C) modes, and the BLE beacon.

## 7. HASITE: Hardware and implemented ITNs

It is first important to note that the selection of hardware components mentioned in this Section was aimed at minimizing the total cost of the system and the power consumption, whilst covering different basic functionalities of a home automation system.

HASITE's subsystems were implemented by using the following components:

- Control and management subsystem. This subsystem made use of a Raspberry Pi 2 Model B, which has an ARM-Cortex A7 quad-core processor running at 900 MHz, and 1 GB of RAM. The operating system used was Raspbian, a Linux distribution based on Debian adapted to run on the Raspberry Pi.
- Communications subsystem. The ITN connectivity is provided by an ESP8266, a Wi-Fi module based on a low cost SoC that supports IEEE 802.11 b/g/n communications at 2.4 GHz, Wi-Fi

Protected Access (WPA/WPA2), and IPv4. The power consumption of these modules is very low: 10 *µ*A in deep sleep mode, and between 70 mA and 240 mA in active mode. Although the ESP8266 offers different operating modes and communication protocols, the ITNs made use of AT Commands through Serial TTL.
- Transducers subsystem. To test HASITE's features, several sensors and actuators were selected considering the most common sensing and actuation needs in home automation appliances. The selected sensors were the following: a basic low-cost digital temperature and humidity sensor (DHT22), a digital ambient light sensor (BH1750FVI), a PIR motion sensor module (HC-SR501), and a voltage and current sensor (INA219). A couple of actuators were also tested: a SRD-05VDC-SL-C relay and a light dimmer based on a TRIAC TIC206 and optocouplers.
- ITNs. Each ITN receives and processes the data either using an ATmega328P or an ATmega32U4 microcontroller, which operate at 16 MHz, provide 14 digital inputs/outputs and 6 analog inputs, and have 32 KB of Flash memory, 2 KB of RAM and 1 KB of EEPROM. They support the acquisition of analog and digital signals via protocols like SPI, I2C (TWI), TTL serial communications or PWM, allowing for the integration of regular COTS inexpensive sensors and actuators in our ITNs, without needing further development or support by manufacturers and integrators. A single ITN with only one microcontroller may have several transducers and can be defined by only one ITE.
- Gateways/Wi-Fi APs. The connections among the different Wi-Fi devices were tested while using two different APs: the main Access Point (AP) was a Linksys WRT54G V2.2 router with a *DD-WRT v24-sp2 (08/07/10) micro (SVN revision 14896)*, while an SMCWEB-N with a *1.0.2.6 (July 6th, 2007)* firmware version was used as secondary AP.

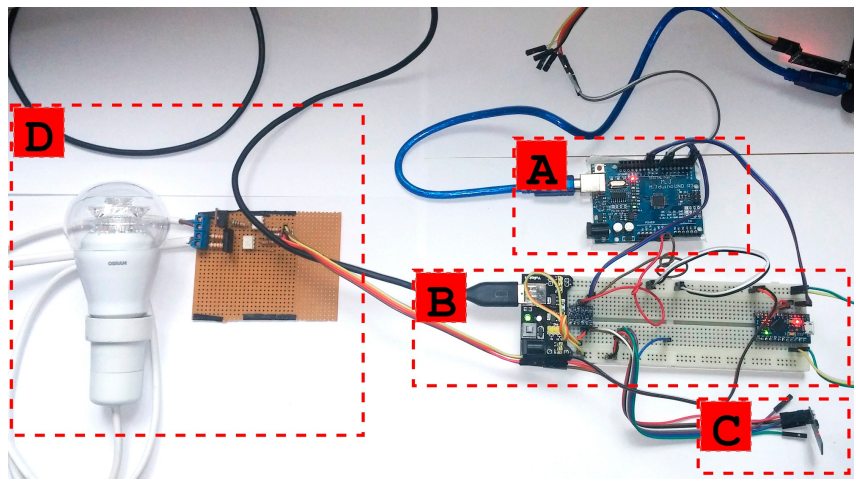

**Figure 9.** Example of a dimming node.

Figure 9 shows, for illustration purposes, the disassembled version of an ITN prototype. It can be observed that the node is composed by an ATmega328P microcontroller (A), a breadboard (B), a power supply module (on the left in B), an ATmega32U4 microcontroller for the dimming driver control (on the right in B), an ESP8266 module (C), and a dimming driver for an adjustable LED bulb (D). All these components can be miniaturized and integrated into a household power socket. The miniaturized result, still in a prototype form factor, is shown in Figure 10. Due to the smaller form factor, the socket version of the ITN replaces the ATmega328P of the original prototype in Figure 9 with an ATmega32U4. In addition, the breadboard and the power supply module were replaced by direct connections and a USB mobile phone charger, respectively.

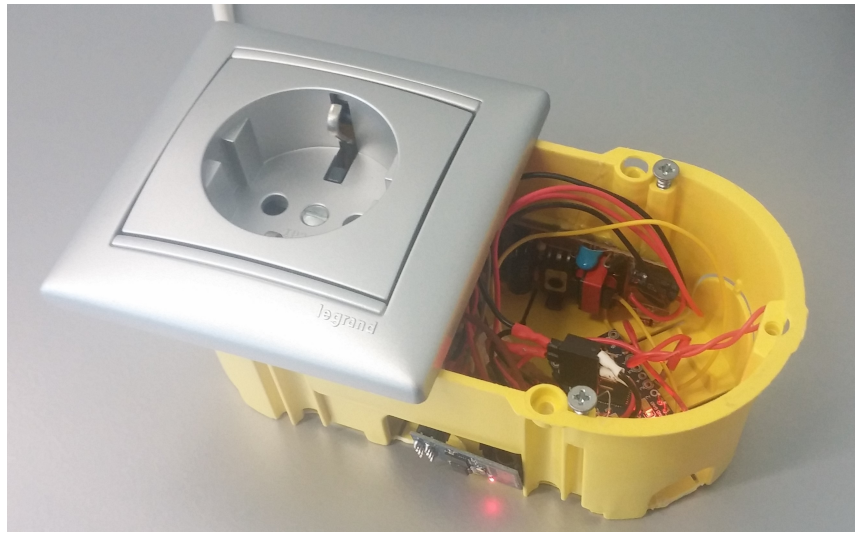

**Figure 10.** Light dimming ITN integrated into a wall power socket box.

## 8. Results

In this Section we present the results of several tests conducted to evaluate the performance of HASITE in real scenarios. Two different sets of tests were conducted. The first set of tests was aimed at measuring the time that went by since an ITN is plugged until it is accessible to an end user. The test was designed to isolate easily the times depending on the self-configuration and self-registration protocols from the ones dependent on the physical and data link layers (in this case, Wi-Fi). The second set of tests was aimed at determining whether HASITE's hardware and software were fast enough to satisfy the response times and processing capabilities desired in a home automation application.

In order to perform the different tests, we modified an ITN to register several relevant parameters (for instance, the time spent at each of the steps of the self-configuring and self-registration protocols). In this way, the ITN was programmed to record and dump the results to a computer via a serial TTL communication.

The tests were conducted inside GTEC's Lab, in the Scientific Area of the University of A Coruña. Figure 11 shows the testing environment, where L denotes the position of the laptop that ran the tests and R was the location of the Wi-Fi router used as access point. The ITN and the laptop were connected via Wi-Fi to the router. Such a router was placed at a height of 80 cm above the floor, while the ITN was moved through the building and positioned in four different relevant locations:

- Scenario A (short distance, line-of-sight): the ITN was located 1 m from the router at same height.
- Scenario B (medium distance, line-of-sight): the ITN was located 5 m from the router at same height.
- Scenario C (medium distance, non-line-of-sight): the ITN was located 8 m from the router at the same height with a closed plywood door in between.
- Scenario D (medium distance, non-line-of-sight): the ITN was located 10 m from the router, upstairs and with both a closed door and stairwells in between.

Such scenarios were selected to have a set of diverse data samples, minimizing any bias introduced by the distance between the AP and the ITN, or by interferences from any Wi-Fi network operating in a frequency close to the one used in the tests.

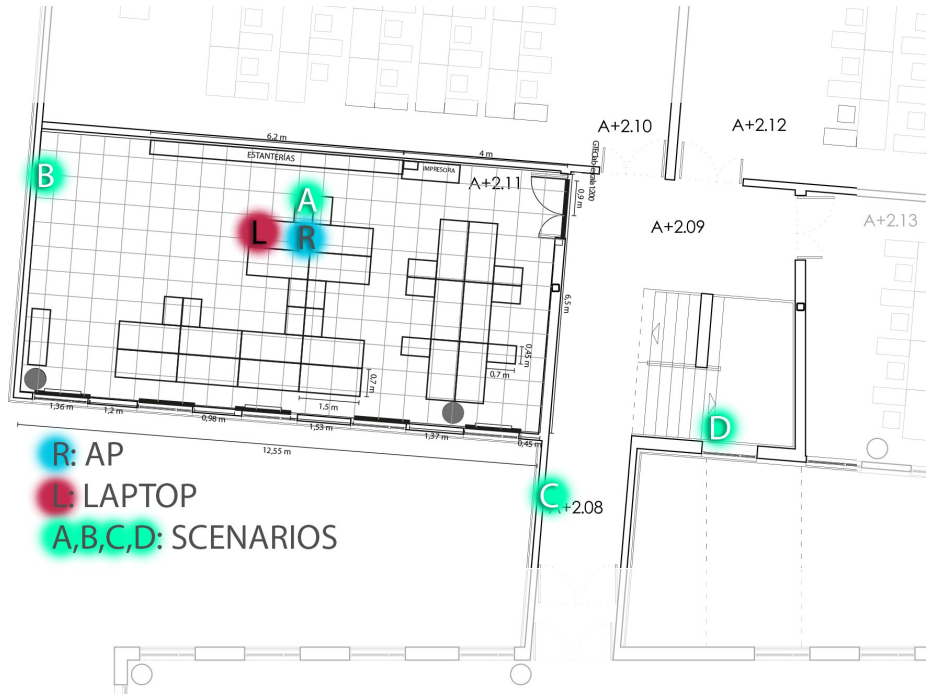

**Figure 11.** Testing environment.

*8.1. Self-configuration and Self-registration Time*

The first test was performed to estimate the time spent at each of the states shown in Figure 5. The objective consisted in determining possible bottlenecks. We performed 20 independent realizations in each of the four scenarios described in the previous subsection.

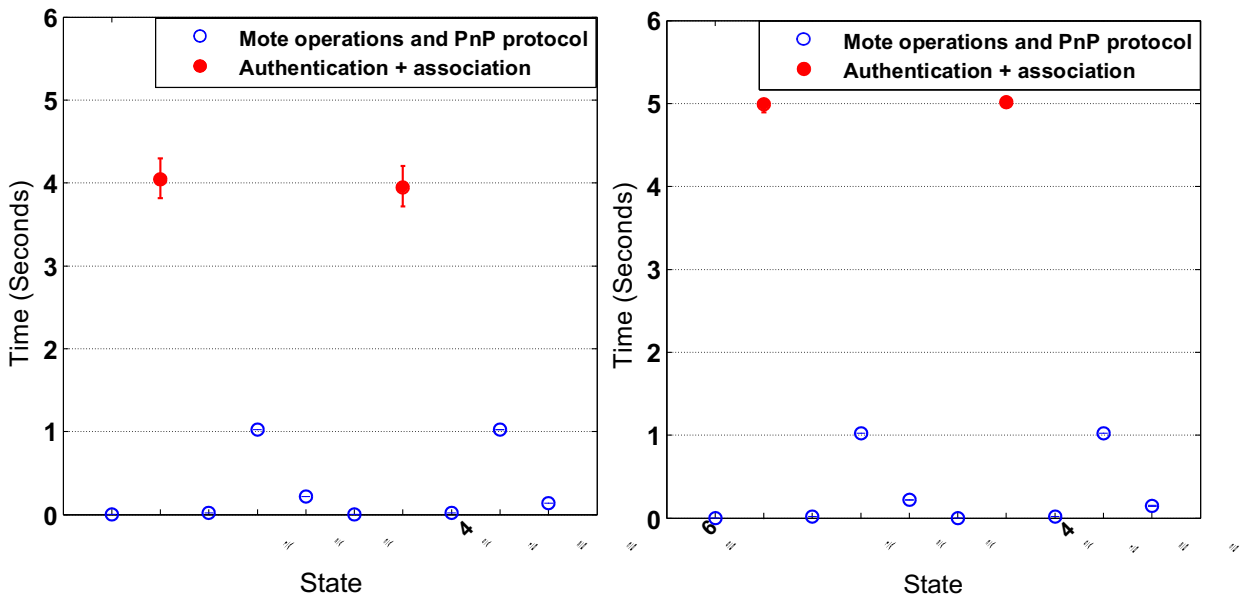

**Figure 12.** Self-configuration and self-registration times using Linksys WRT54G AP (left) and SMCWEB-N AP (right).

In order to evaluate the impact of the AP on the whole plug-and-play process, two different devices were tested: a Linksys WRT54G and an SMCWEB-N. Figure 12 shows the time needed to perform the self-configuration and the self-registration protocols (the x-axis labels correspond to the different protocol states explained in Subsection 5.2). This Figure shows the central point and the confidence intervals with a confidence level of 95%. It can be seen that most of the time is spent on the authentication and the association between the AP and the ESP8266 (shown in red).

Table 3. Network packet analysis of an ID request to an ITN.

| Frame Number | Time (seconds) | Source | Destination | Protocol | Length (bytes) |
|---|---|---|---|---|---|
| 96 | 5.329233000 | 192.168.1.119 | 192.168.1.123 | HTTP | 127 |
| 97 | 5.378788000 | 192.168.1.123 | 192.168.1.119 | TCP | 123 |
| 98 | 5.429015000 | 192.168.1.119 | 192.168.1.123 | TCP | 54 |
| 99 | 5.436950000 | 192.168.1.123 | 192.168.1.119 | HTTP | 60 |
| 100 | 5.437021000 | 192.168.1.119 | 192.168.1.123 | TCP | 54 |
| 101 | 5.437993000 | 192.168.1.119 | 192.168.1.123 | TCP | 54 |
| 102 | 5.444188000 | 192.168.1.123 | 192.168.1.119 | TCP | 60 |
| 103 | 5.489354000 | 192.168.1.119 | 192.168.1.123 | TCP | 66 |
| 104 | 5.493145000 | 192.168.1.123 | 192.168.1.119 | TCP | 60 |
| 105 | 5.493232000 | 192.168.1.119 | 192.168.1.123 | TCP | 54 |

Analyzing the results obtained for the Linksys WRT54G (on the left in Figure 12, it can be concluded that the confidence intervals corresponding to the authentication and association steps are the most variable ones. Both the self-configuration (states 4 and 5-A) and self-registration protocols (states 6 and 5-B) are completed in less than 2 s. Considering all the states, the average time of all the 80 realizations was 10.4412 s with a variance of 1.8241 s.

The results obtained using the SMCWEB-N AP (on the right in Figure 12) show that the authentication and the association times are larger than the ones obtained with the Linksys WRT54G, but they oscillate less. With the SMCWEB-N, the average time of the whole process was 12.4871 s and the variance 0.2571 s.

After observing the results of both APs it can be concluded that both self-configuration and self-registration protocols play a minor part on the times of the whole process, being the time of actually connecting to the AP the most relevant one. It also can be seen that the selection of an AP directly affects the total time and variance. In both cases, the time required for the entire process was less than 13 seconds, measured from the time the ITN is connected to the power supply, until it is available to be accessed through the network.

*8.2. ITN Response Time*

The response time of a node is defined as the time that goes by since a request to access a sensor value is issued until it is finally performed and the response is sent by the ITN. In order to simulate this behavior, the time response was measured using a script in Python, which requests each node ID periodically for a number of times. This replicates the sequences of processes required to access a sensor value or to change a transducer state. The script issues a request to an ITN, and the ITN sends back a response containing a JSON object with its ID (e.g. {"NodeID":"7.1.1"}). Using this method, the script was run 1 000 times for each scenario using the Linksys WRT54G router. A wait time of 25 ms was included between requests to avoid overloading the ITN. At the same time, it was captured the Wi-Fi load of the router using its web interface (as shown in Figure 14).

Moreover, the script counts the number of successful and failed requests, and measures the time required to perform each one. For all the tests carried out during 1 000 script executions, a 0% ratio of failed requests was obtained. In a longer test that executed the script 200 000 times, there were only 4 failed requests, meaning an error ratio of 0.002%.

Table 3 shows different results obtained with the network analyzing tool Wireshark at scenario A. The following parameters are included: frame number, time instant of the request, source and destination IP addresses (the ITN has IP 192.168.1.123, whilst the client has IP 192.168.1.119), the protocol, and the length of each packet captured. Considering the length of all packets, we have obtained that the whole communication uses 712 bytes of data.

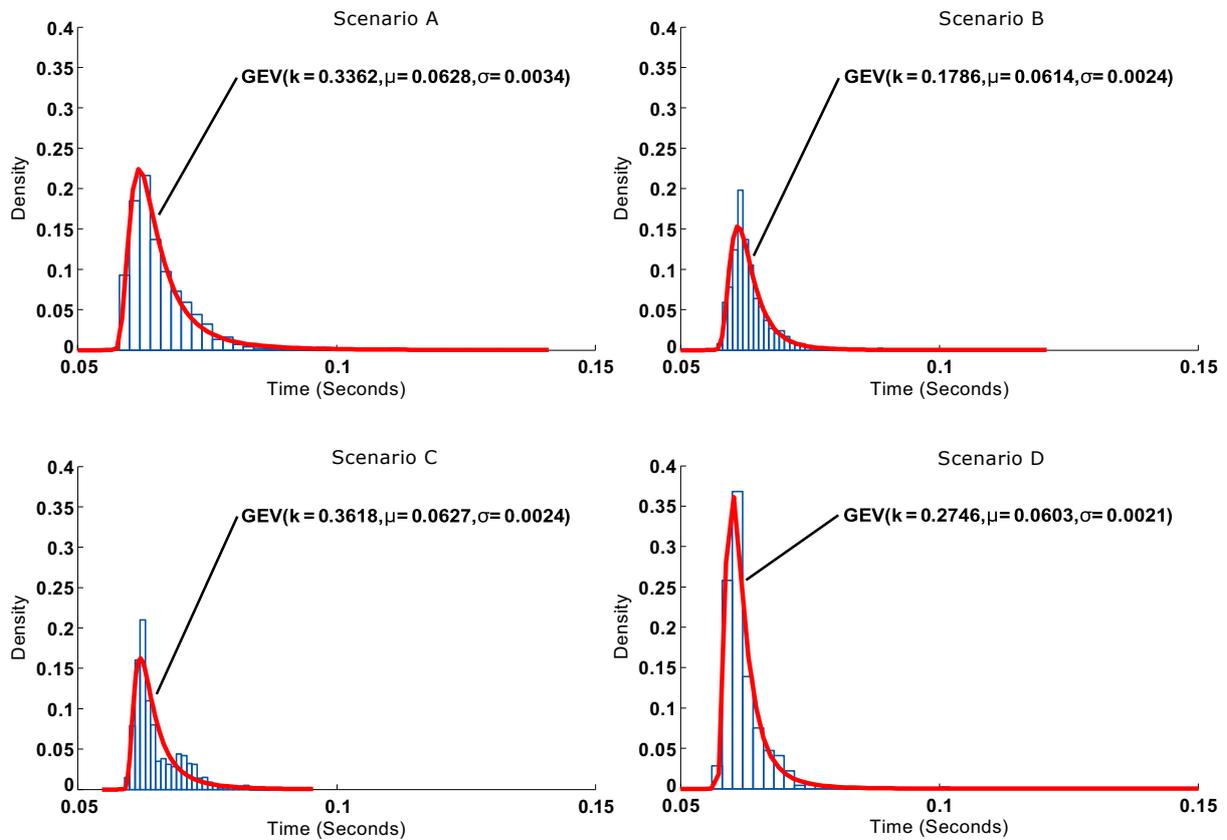

**Figure 13.** Response time for the scenarios considered in the test.

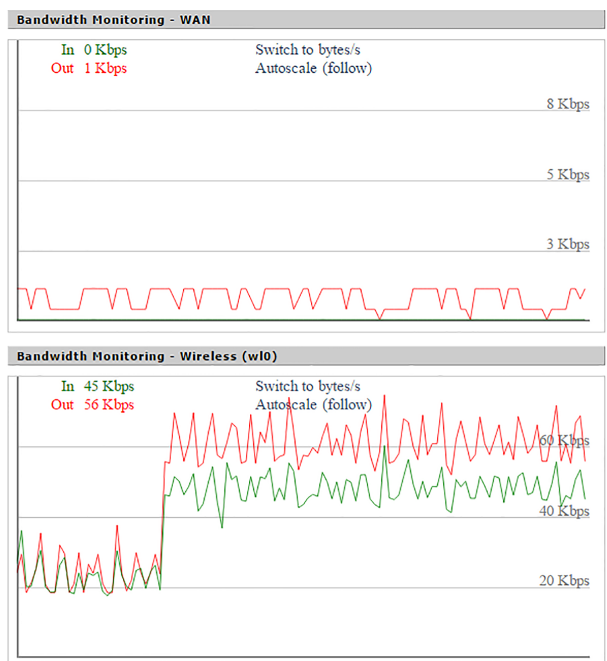

**Figure 14.** Access Point load evolution during the test.

The histogram corresponding to the response time for the four scenarios is depicted in Figure 13. As it can be seen, in the selected scenarios, the response time is not directly related to the distance between the node and the AP. In addition, for each scenario, the figure shows a curve that fits almost perfectly the histogram: a Generalized Extreme Value (GEV) distribution with its corresponding parameters of location ($\mu$), scale ($\sigma$) and shape ($k$) [45]. It is important to note that such curves can be used to model the behavior of the nodes and then generate artificial samples from the fitted distributions to perform Monte Carlo simulations.

The location and scale of the distributions show that the best results are obtained in Scenario D, while Scenario A yields larger times. This can be explained due to the fact that in Scenario A the AP, the node, and the laptop used to monitor the process were placed very close to each other (i.e. they were placed within a 1 m radius). In Scenario D, the AP and the ITN were placed further, but they had almost direct line-of-sight, only obstructed by a plywood door that attenuates slightly the signal.

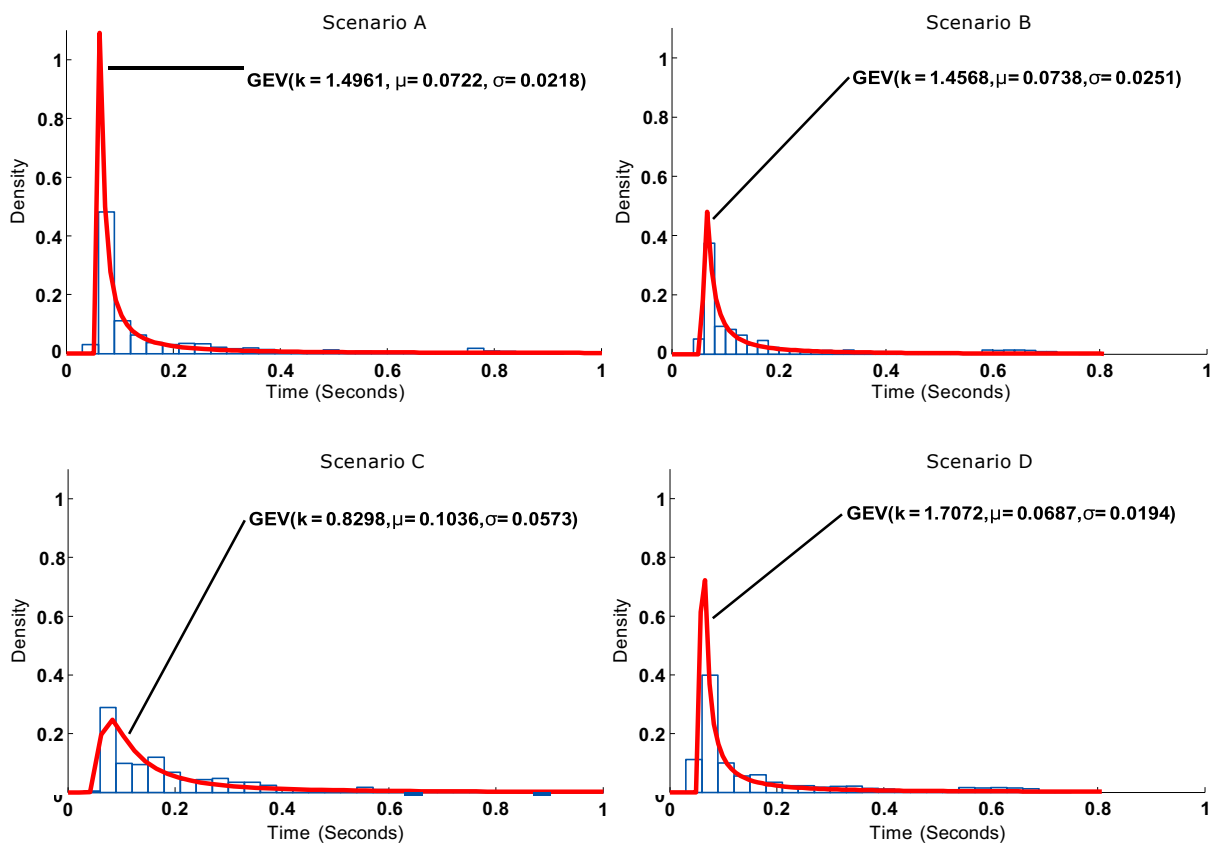

**Figure 15.** Response time with heavy load on the Wi-Fi segment obtained for the scenarios considered in the test.

In general, it can be concluded that the differences between the scenarios are minimal. For instance, note that the time difference between the means of the worst and best scenarios is 4.4 ms (i.e., a difference of less than 7%). The response time is below 70 ms in all scenarios. Also note that, although the tests were performed in real scenarios, where a significant number of nearby Wi-Fi APs were transmitting simultaneously, the results allow us to state that distance is only relevant in terms of connectivity range.

Another test was performed in order to simulate some load on the Wi-Fi network. For that, a big dummy file was downloaded to a computer connected to the same AP as the node. Figure 15 shows the histogram obtained for the different scenarios. Comparing Figure 13 and Figure 15, an increase in response time can be seen, but it remains below 220 ms.

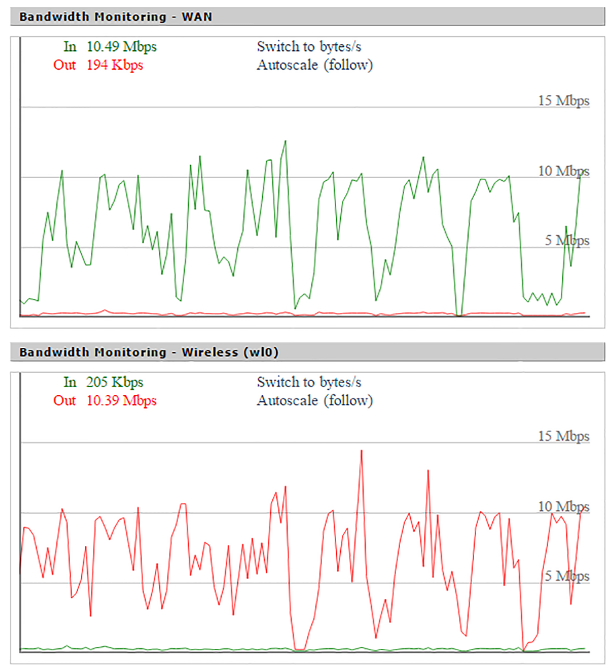

**Figure 16.** Access Point load evolution during the test with heavy load on the Wi-Fi segment.

Taking a closer look at the AP load in Figure 16 and comparing it with Figure 14, abrupt speed drops can be observed. This is due to the heavy load artificially added with the laptop, what makes the AP load go over 1.0 (more than the router can handle) producing those speed drops and making the confidence interval grow larger than in the no-load test, thus producing larger response times. These results allow us to conclude that, even in a situation where the Wi-Fi network was overloaded, the delay perceived by a user would be minimal and the system would be completely functional.

Regarding the distribution and fitting curves of Figure 15, the results give more information about the transfer rate. Again, Scenario D is the one with the smallest times. However, with load, Scenario C is not only the worst, but it differs greatly from the three other scenarios (the parameters location and scale have much greater values at this spot). By observing the map represented in Figure 11 it can be seen that Scenario C has a wall directly in-between the ITN and the AP, what can cause the observed drop in performance. Moreover, as explained before, the large number of APs on the building can interfere and derive into relevant variations on performance with small variations of location. Figure 17 summarizes the ITN interaction response time tests for the four scenarios in the event of both heavy and no load.

## 9. Conclusions

This paper presented the design and implementation of HASITE's architecture, which is focused on a plug-and-play mechanism for home automation systems. The main elements of HASITE are a Home Automation Gateway, Intelligent Transducer Nodes (ITN) and Wi-Fi Access Points (AP). In order to control these elements, three subsystems were defined: the control and management subsystem, the communications subsystem and the transducers subsystem. Intelligent Transducer Enabler (ITE) data structures are a key element in HASITE infrastructure, and are aimed at storing configuration data, easing the access, and at communicating and interpreting the data managed by each ITN and its transducers.

The self-configuring protocol proposed allows the ITNs to be configured automatically. It relies on a configuration AP and a series of steps between the Home Automation Gateway and the ITNs to request and share the configuration information. This mechanism has three different modes of operation: automatic, dynamic request and previous identifier registration. Each of these modes of operation is aimed at easing the addition of ITNs and providing different levels of security.

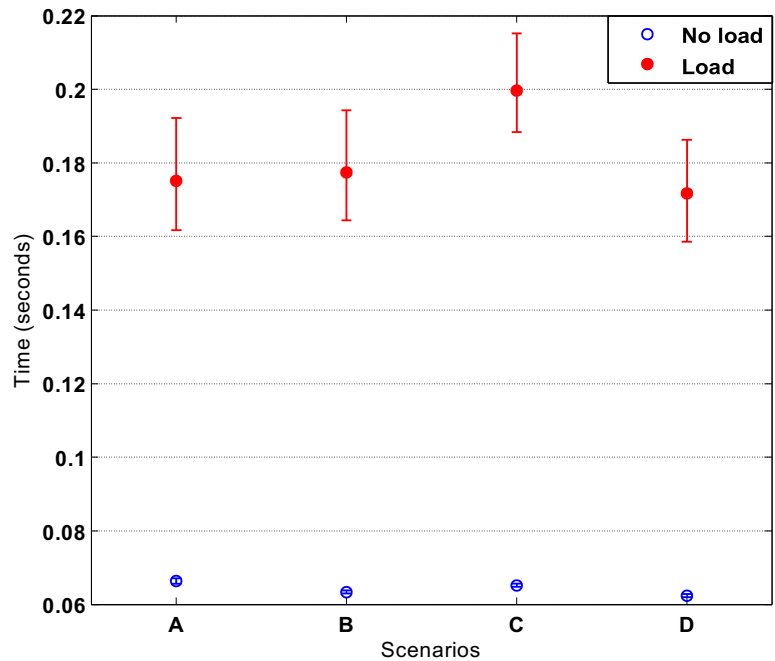

**Figure 17.** Integration of all ITN interaction response time tests.

　　The novel self-detection protocol is able to detect an ITN once configured, storing the record of each node registered, and making it available for further access and interaction. The system has an API to access the basic configuration parameters, the registered ITNs, the information of the ITEs, and the inputs and outputs of the transducers. Furthermore, a hybrid application for Android and iOS was developed using the API. The application made use of BLE beacons to estimate the user location within a house. Using the approximate location of the user, the application can adjust and customize the interface to display only the controls and information needed for controlling nearby appliances. This smartphone application reflects the power of the ITEs and the API access: all registered ITNs and the ITEs can be obtained with simple GET requests. Once obtained, the information contained on an ITE allows for the easy understanding of the different sensors and actuators controlled by an ITN. The messages, units, maximum and minimum values, among other useful information, can be easily understood by a human, thanks to the use of JSON. In addition, an ITE can be parsed to assign software or hardware controls and displays to a transducer automatically. This simplifies not only the communication with HASITE and its ITNs, but also the creation of reusable GUI elements for similar transducers.

　　The use of ITEs as a leading element grants full support of the IoT paradigm, allowing for the use of any RDF notation, such as JSON-LD, and the creation of an ontology around the transducers, what would add automatic control and improve decision making.

　　HASITE's architecture was tested using a low-cost and low-consumption SBC solution as Home Automation Gateway. Several common transducers (a relay, a light dimmer, and humidity, temperature, presence and current sensors) were embedded into ITNs. The time performance of the implemented protocols was determined in four different locations of a building. The system was evaluated while using two different APs and different load conditions on the Wi-Fi network. The results were encouraging: the total time was always less than 13 seconds, being the self-configuration and self-registration protocols completed in less than 2 seconds.

**Acknowledgments:** This work has been funded by the Spanish Ministry of Economy and Competitiveness under grants TEC2013-47141-C4-1-R and TEC2015-69648-REDC.

**Abbreviations**

The following abbreviations are used in this manuscript:

API: Application Programming Interface
BACnet: Building Automation and Control Networks
DOMOSEC: Domotics and Security
EIB: European Installation Bus
GPIO: General-purpose input/output
GPRS: General Packet Radio Service
GSM: Global System for Mobile
HASITE: Home Automation System based on Intelligent Transducer Enablers
HTTP: Hypertext Transfer Protocol
I2C: Inter-Integrated Circuit
IoT: Internet of Things
IP address: Internet Protocol address
ITE: Intelligent Transducer Enabler
ITN: Intelligent Transducer Node
JSON: JavaScript Object Notation
LabVIEW: Laboratory Virtual Instrumentation Engineering Workbench
LonWorks: Local Operating Network
RAM: Random Access Memory
REST: REpresentational State Transfer
RF: Radio Frequency
RFID: Radio Frequency IDentification
SBC: Single Board Computer
SCITS: Self-Configuring Intelligent Transducer System
SPI: Serial Peripheral Interface
TEDS: Transducer Electronic Data Sheets
UART: Universal Asynchronous Receiver-Transmitter
UPB: Universal Powerline Bus
URI: Uniform Resource Identifier
URL: Uniform Resource Locator
WPA: Wi-Fi Protected Access